
\defenva{proposition}{\bf }{\it }{Proposition}
\def\R{{\Bbb R}}
\def\Spin{{\Bbb S}}
\def\C{{\Bbb C}}
\def\omicron{{o}}
\def\whst{(\widehat M,{\widehat g}_{ab} )}
\def\st{{(M,g_{ab})}}
\def\whSig{{\widehat{\Sigma}}}
\def\cN{{\cal N}}
\def\cH{{\cal H}}
\def\scri{{\cal I}}
\def\scrip{{{\cal I}^+}}
\def\Int{\mathop{\rm Int}}
\newcount\EEK
\EEK=0
\def\eek{\global\advance\EEK by 1\eqno(\the\EEK )}
\MAINTITLE{A Phase Space for Gravitational Radiation}
\AUTHOR{Adam D. Helfer}
\INSTITUTE{Department of Mathematics, University of Missouri, Columbia,
MO 65211}
\DATE{}
\SUMMARY{We give a new definition, based on considerations of well-posedness
for a certain asymptotic initial value problem, of the phase space for
the radiative degrees of freedom of the gravitational field in exact General
Relativity.  This space fibres over the space of final states, with
the fibres being the purely radiative degrees of freedom.  The symplectic form
is rigorously identified.

The infrared sectors are shown to be the level surfaces of a moment map of an
action of the quotient group Supertranslations/Translations.  A similar result
holds for Electromagnetism in Minkowski space.}
\maketitle

\titlea{Introduction}

The theory of gravitational radiation interweaves physical and mathematical
progress to an unusual degree:  some physical intuition is necessary to begin
to pose questions which are significant; and the construction of a
mathematically satisfactory theory often serves as a test of that intuition.
If we define radiation mathematically as disturbances which escape to (or come
in from) infinity, then gravitational radiation, which travels at the speed of
light, will escape to null infinity, and ought to be analyzed there.  Since
what ``null'' is is determined by the dynamical field, from the analytic point
of view one is studying the asymptotics along characteristics of a system of
quasilinear equations.
There are several, inequivalent, mathematical meanings this might have, and
the nature of the physical questions
under consideration must guide the analysis.

It was Bondi and coworkers (Bondi 1960, Bondi, van der Burg and Metzner 1962,
Sachs 1962a; see also Newman and Penrose 1962) who introduced the idea of
analyzing the field at large null separations, and Penrose (1963) who recast
their asymptotic conditions as the existence of a null hypersurface $\scri$ at
null infinity.  After this work, it was clear that at a formal level the
radiative degrees of freedom were represented by a function, now called the
Bondi shear, on $\scri$.

Ashtekar \& Streubel (1981) seem to have been the first to realize that it
would
be desirable to develop, not just a space-time by space-time analysis, but a
phase space of radiative degrees of freedom in terms of the asymptotics of the
fields.  They gave such a construction of a phase space entirely in
terms of (functions equivalent to) the Bondi shear.  This resulted in a
new derivation of the Bondi energy-loss formula, and a new expression for the
total angular momentum emitted in gravitational waves (which has since been
confirmed by independent means; see Helfer 1990).  Also Ashtekar (1987)
attempted to use this phase space as the basis for a construction, by geometric
quantization, of an $S$-matrix theory of gravitons.

It is not straightforward to decide whether such a phase space is correct.  The
strongest argument would be to show that it was an appropriate limit of phase
spaces constructed from data on Cauchy surfaces.  A program for such a proof
was outlined by Ashtekar \& Magnon-Ashtekar (1982); their idea was to deform a
Cauchy surface to $\scri$.  They were able to show, at a formal level, that the
symplectic form of Ashtekar \& Streubel was correct.  However, they also
pointed out that there were some analytic limitations which might preclude the
conclusions of the formal argument from being valid generally.  It seemed
likely
that in many cases -- perhaps generically -- the symplectic
form would have to be supplemented by ``leakage'' terms at timelike infinity.
Also, if one considers a space-time which radiates in such a way as to become
vacuum to the future of some asymptotically null hypersurface, then the sorts
of perturbations allowed by the Ashtekar-Streubel phase space will include
space-times with negative Bondi-Sachs energies, which one would like to rule
out as unphysical in light of the positivity theorems (Ludvigsen \& Vickers
1981, Horowitz \& Tod 1982, Schoen \& Yau 1982).  So the successes of the
Ashtekar-Streubel construction suggested that it ought to be (at least largely)
correct, but it seemed difficult to justify rigorously.

I shall construct a phase space of the radiative degrees of freedom
of the gravitational field in exact General Relativity.  The isolation of these
degrees of freedom is determined by considerations of well-posedness for a
certain asymptotic initial value problem (the ``M-shaped'' hypersurface
problem, which is well-posed by recent theorems of Friedrich 1983, 1992 and
Rendall 1990, 1992.)
We shall be able to make precise the sense in which the
Ashtekar-Streubel symplectic form is valid, and understand why there are no
leakage terms.

The most important consequence is a shift in the idea of
what a radiation problem is.  In order to isolate ``all'' of the outgoing
radiative degrees of freedom, one must take two limits:  one must move out to
$\scri$; and one must also consider arbitrarily late retarded times.  Although
for any sufficiently well-behaved space-time one can consider both limits, one
cannot reverse this, and solely from the radiative data construct the
space-time.  One must give the internal state at some late retarded time,
together with the radiation emitted prior.\fonote{It may at first be more
intuitive to consider the time-reversed case, where one specifies a space-time
by giving the internal state at some advanced time, together with the
radiation incoming thereafter.}  From this point of view, the
Ashtekar-Streubel phase space is an approximation in which radiation has
wholly decoupled from the internal degrees of freedom.  The coupling which
does exist is rather mild (since the radiative data must match up to the
internal data only at one cut of $\scri$), so the approximation is valid
for many purposes.

{}From this point of view, the total phase space has the structure of a fibre
bundle, with the radiative degrees of freedom (prior to the late retarded
time) fibering over the space of final states (at the late retarded time).
This is because the radiative data must agree with the internal data at the
cut of $\scri$ corresponding to the retarded time.  To
a large extent, the fibres have simple structures, and most of the
complications are in the base space.  (This is related to the fact that
$\scri$ is a null hypersurface, and so data on it are unconstrained.)  We are
concerned here only with a fibrewise analysis.  An understanding of the total
phase space would be a natural next step; one would expect a rigorous analysis
of the linearization stability of the ``M--shaped'' problem to be possible (cf.
Moncrieff 1975, 1976).

We consider a somewhat different class of radiative data from those of
previous constructions.  One difference is quite minor (we work with a
generalization of test functions, rather than a weighted Fr\'echet space) and
not expected to be physically significant, but another is important:  we allow
``infrared'' fields.  These are fields which, although pure gauge at early and
late retarded times, nevertheless have a gauge mismatch between these two
regimes.  We shall find that the residual gauge freedom at $\scri$ gives rise
to a foliation on the phase space which turns out to be a classical analog of
the emergence of infrared sectors in Quantum Electrodynamics.  The discovery
of these was a surprise in quantum theory, but we can now see that it is
presaged (logically, if not historically) by the foliation of the phase space.

The present construction seems likely to help with a similar infrared problem
in
the asymptotic quantization program for General Relativity.  In Quantum
Electrodynamics, the infrared sector for the outgoing electromagnetic field is
determined by the asymptotic state of the outgoing charged particles, and this
identification is important in constructing a separable Hilbert space.  It had
seemed that there would be no way of deciding on the infrared sector for the
outgoing gravitational waves (Ashtekar \& Narain 1981, Ashtekar 1987).
However,
it now seems that the internal state of the field at a late retarded time could
determine the infrared sector.  This will be considered elsewhere.

Finally, in a separate paper, I shall show how this phase space can be used to
derive the asymptotic motions of space-time to which the angular momenta at
$\scri$, as defined by Penrose (1982), are conjugate.  In this analysis, not
only is knowledge of some of the internal data necessary, but also the
foliation of the phase space by the infrared sectors enters crucially.

The plan of the paper is this.  The next Section is devoted to a review of the
definitions which will be needed:  spin- and boost-weighted functions; Bondi
coordinates; the Bondi-Metzner-Sachs group; and the reduction of a phase space
by an abelian gauge group.  In Section 3, we sketch the application of our
ideas in the case of electromagnetism, and show how the infrared sectors
emerge.  Section 4 reviews the results of Friedrich and Rendall, and
establishes the existence of maximal globally hyperbolic solutions of the
initial-value problem (Theorem ~4.1).  In Section 5, the manifold underlying
the
phase space is constructed as a certain function space.  In Section 6, the
main technical result, the correctness of the Ashtekar-Streubel symplectic
form is established (Theorem~6.1).  It is perhaps surprising that such a
rigorous result is presently possible, since a direct approach to this problem
would require better asymptotic estimates for Einstein's equations than are
known.
The last Section discusses the infrared
sectors for gravity.  These are related to motions of the phase space in a
somewhat different fashion from those for electromagnetism.

We treat only the case of pure gravity, but it should be evident that coupling
to matter can be accomodated within this framework.

\it Conventions.  \rm
Our conventions for space-time quantities are those of Penrose \& Rindler
(1984-6).  We will be concerned with calculus on the phase space and on the
space-time.  It will be convenient to use the standard spin-coefficient
formalism on the space-time, and the modern coordinate-free notation for the
phase space.  We use $\langle s,\lambda\rangle$ to stand for the duality
pairing between a vector $s$ and a covector $\lambda$ in this case.  Thus
the symbol $d$ represents a gradient in the phase space except when it occurs
in a line, area or volume element, $du$, $d{\cal S}$ or $d{\cal S}\, du$, with
respect to the Bondi coordinate system on $\scri$.

\it Choice of Function Spaces.  \rm
In most circumstances, the particular choices of function space used in General
Relativity are not of physical significance, and are made for mathematical
convenience.  The real question is whether the choice at hand admits a
sufficiently broad class of fields to be of interest.  For the analysis of
gravitational radiation, although the formalism of null infinity was
introduced in the early sixties, it is only recently that existence theorems
for
solutions to the Einstein equations with the right sort of control of the
asymptotics have begun to be proved.

As of this writing, it is not clear whether there is a physically preferred
degree of smoothness for the asymptotic regimes.  We have chosen to work in
the $C^\infty$ category, since the theorems of Friedrich and Rendall make this
possible and in the absence of any other choice it seems the most natural.
The function spaces we use are analogs of the
familiar spaces of test functions, and so are unphysical approximations in
that they become exactly zero (or pure gauge) in various regimes where a
realistic field would only decay.  However, it will be apparent that our
constructions could accomodate such behavior, and fields of low
differentiability, without difficulty.

\titlea{Preliminaries}

We collect here some terminology which will be assumed in what follows.

\titleb{Spin-  and Boost-Weighted Functions}

The radiative degrees of freedom of the graviational field will be encoded in
a function called the Bondi shear.  This function takes values in a certain
complex line bundle over null infinity; in classical language, the function
has spin and boost weight.  We here recall the definitions of these line
bundles.

Let $\Spin ^A$ be spin space.  It is a two-complex dimensional vector space,
and its associated projective space is identified with the Riemann sphere $S^2$
equipped with a conformal structure and an orientation, but not a metric.  To
accord with standard conventions, we shall use $\omicron ^A$ for a general
(non-zero) element of $\Spin ^A$; the class of spinors proportional to
$\omicron ^A$ defines a point on the sphere.  We may sometimes introduce a unit
future-pointing timelike vector $t^a$.  The subgroup of ${\rm SL}(2,\C )$ which
preserves this vector is ${\rm SU}(2)$ (in a basis in which the spinor
$t^{AA'}$
is diagonal), and so $t^a$ defines a unit sphere metric on $S^2$.  We put
$\iota ^A=t^{AA'}\omicron _{A'}/(t^{BB'}\omicron _B\omicron _{B'})$;
then $\omicron _A\iota ^A=1$.

A function on $\Spin ^A$ which may not be holomorphic is conventionally written
$f(\omicron ^A,{\overline\omicron}^{A'})$.  The function is said to be \it of
type $\{ p,q\}$ \rm if
$$f(\lambda\omicron ^A,{\overline\lambda}{\overline\omicron}^{A'})
  ={\strut\lambda}^p{\strut\overline\lambda}^qf(\omicron
^A,{\overline\omicron}^{A'})\eek$$
on $\Spin ^A-\{ 0\}$.
(One says $f$ has \it spin-weight \rm $(p-q)/2$ and \it boost-weight
\rm $(p+q)/2$.)  Such a function is a section of a certain complex line-bundle
over the sphere.  The line-bundle will be denoted simply as $\{ p,q\}$.
Its space of smooth sections will be denoted $C^\infty (S^2,\{ p,q\} )$.

There are two important differential operators on these line bundles:
$$\eth :C^\infty (S^2,\{ p,q\} )\to C^\infty (S^2,\{ p+1,q-1\})\eek$$
defined by
$$\eth f=\sqrt{2}t_{AA'}\omicron ^A{\overline\omicron}^{A'}
 \overline{\iota}^{B'}{ {\partial\phantom{ {\overline\omicron}^{B'}} }
  \over{\partial {\overline\omicron}^{B'}}}f\eek$$
and its conjugate $\eth ':C^\infty (S^2,\{ p,q\})\to C^\infty (S^2,\{
p-1,q+1\})$.

Notice that the conjugate of $\{ p,q\}$ is $\{ q,p\}$.  In particular,
elements of $\{ p,p\}$ have unique ${\rm SL}(2,\C )$-invariant decompositions
into real and imaginary parts.  There is also a hidden ${\rm SL}(2,\C
)$-invariant Hermitian structure for certain other spaces of sections of the
line bundles.  It turns out that the operator $\eth ^{q+1}:C^\infty (S^2,\{
p,q\})\to C^\infty (S^2,\{ p+q+1,-1\} )$ is ${\rm SL}(2,\C )$-invariant.  Let
us
consider the case $p=q$.  Then this operator is surjective and its kernel is
the complexification of a real ${\rm SL}(2,\C )$-invariant subspace of
$C^\infty
(S^2,\{ p,q\})$.  This allows us to write each element in the image as a sum of
the image of a real section and the image of an imaginary section.  These are
the \it electric \rm and \it magnetic \rm parts of $C^\infty (S^2,\{ 2q+1,-1\}
)$ (Newman \& Penrose 1966).   The important case of this decomposition for us
will be for the Bondi shear, $\sigma\in C^\infty (S^2,\{ 3,-1\})$.  We can
always write $\sigma =\eth ^2\lambda$ for some $\lambda \in C^\infty (S^2,\{
1,1\})$.  The electric part of $\sigma$ is $\eth ^2\Re\lambda$ and the magnetic
part of $\sigma$ is $\I\eth ^2\Im\lambda$.

\titleb{Bondi Coordinates}

Bondi coordinates are appropriate for radiation problems.  They are defined as
follows.  (See Penrose \& Rindler 1984-6 for a fuller discussion.)

Let $\whst$ be an oriented, time-oriented space-time.  Suppose that ${\widehat
M}$ embeds in a manifold with boundary $M$, and there exists a smooth function
$\Omega$ on $M$, positive on $\widehat M$ and zero on $\scrip =\partial M$,
with
nowhere-vanishing gradient on $\scrip$.  Suppose also $g_{ab}=\Omega ^2
{\widehat g}_{ab}$ extends smoothly to a non-degenerate metric on $M$, and that
all points on $\scrip$ are future end-points of null geodesics.  Then we call
$\scrip$ {\it future null infinity.}\fonote{This differs from some
definitions in that we do not require \it every \rm $\whst$-future endless null
geodesic to have an end-point on $\scrip$.}

Suppose further that $\scrip$ is a $g_{ab}$-null hypersurface, diffeomorphic to
$\R\times S^2$, with the ``$\R$'' factors being its null generators, and that
the field $n_a=-\nabla _a\Omega$ is shear-free on $\scrip$.  Then we can choose
the conformal factor so that cross-sections of the fibration $\scrip\to S^2$
are
unit spheres, and we do so.  We can also choose $n^a$ to be $g_{ab}$-constant
up the generators of $\scrip$, and we do this.  Then a function $u$ satisfying
$n^a\nabla _au=1$ is a \it Bondi retarded time coordinate.  \rm  Fixing such a
function, we let $(\theta ,\varphi )$ be standard polar coordinates on the
cross-sections $u=$ constant, with $n^a\nabla _a\theta =0=n^a\nabla
_a\varphi$.  Then $(u,\theta ,\varphi)$ form a \it Bondi system \rm on
$\scrip$, and $(u,\Omega ,\theta ,\varphi )$ form a \it Bondi system \rm on a
neighborhood of $\scrip$.\fonote{We require $(u,-\Omega ,\theta ,\varphi )$ to
be compatible with the orientation on $\whst$.  The minus sign is because
$\Omega$ decreases as one moves towards $\scrip$ from the interior.}  The
abbreviation
$$d{\cal S}=\sin\theta d\theta d\varphi\eek$$
will be used.

For example, for Minkowski space, we let $(u=t-r,r,\theta ,\varphi )$ be
retarded polar coordinates.  We obtain $\st$ by gluing to this $\{ (u,\rho
,\theta ,\varphi )\mid u\in\R$, $\rho\geq 0$, $(\theta ,\varphi )\in S^2\}$
via the relation $\rho =1/r$, and we take $\Omega =\rho$ near $\rho =0$.

It is often convenient to introduce a complex $g_{ab}$-null tetrad
$(l^a,m^a,{\overline m}^a,n^a)$ at $\scrip$ associated with the Bondi system.
Here $n^a$ is as before, $m^a=2^{-1/2}(\partial _\theta-\I\csc\theta\partial
_\varphi )$, and $l^a$ is the unique future--pointing vector satisfying
$l^an_a=1$, $l^am_a=0$.  There is a spinor dyad $(\omicron ^A,
\iota ^A)$ related by
$$l^a=\omicron ^A{\overline\omicron}^{A'}\, ,\ m^a=\omicron
^A{\overline\iota}^{A'}\, ,\ n^a=\iota ^A{\overline\iota}^{A'}\; .\eek$$
We may therefore speak of spin- and boost-weighted functions on any cut of
$\scrip$.  In fact, since $n^a$ is shear-free, flowing along it preserves
complex structure, so we may regard the bundle $\{ p,q\}$ on any
cut as the pull-back of a bundle (denoted by the same symbol) on $\scrip$.

The \it Bondi shear \rm at $\scrip$ is the shear of the $u=$ constant cuts,
given by
$$\sigma =\omicron ^Am^b\nabla _b\omicron _A\; .\eek$$
It takes values in $\{ 3,-1\}$.  Notice that by its definition, the Bondi
shear depends on the choice of Bondi system.

\titleb{The Bondi-Metzner-Sachs Group}

A Bondi system $(u,\theta ,\varphi )$ will not be unique.  There is a group,
the \it Bondi-Metzner-Sachs (BMS) \rm group, acting on the set of Bondi
systems.  If the null generators of $\scrip$ are complete, that
is, if a Bondi paramter $u$ takes all real values for each $(\theta ,\varphi
)\in S^2$, then there is an associated ``active'' action by diffeomorphisms.
These active motions are in a certain sense the asymptotic symmetries of the
gravitational field, and we shall give an account of them.  Although the
following discussion is phrased in terms of the group action, it applies at
the Lie algebra level even if the generators of $\scrip$ are not complete.

The BMS group is generated by two sorts of motions:  the proper,
orthochronous motions ${\rm Lorentz}={\rm O}(1,3)_{\uparrow +}$, which
act much the same way as they do on half of a light-cone;
and the \it supertranslations, \rm which have the form
$$u\mapsto u+\alpha (\theta ,\varphi
)\, ,\ \theta\mapsto\theta\, ,\ \varphi\mapsto\varphi\; ,\eek$$
where $\alpha$ is an arbitrary smooth function.  The BMS group is a semidirect
product of ${\rm Lorentz}$ and ${\rm Supertranslations}$:
$$0\to{\rm Supertranslations}\to{\rm BMS}\to{\rm Lorentz}\to 0\; .\eek$$

The relation of the BMS group to the connected component of the isometry group
of Minkowksi space, the Poincar\'e group, is important.  The Poincar\'e group
is also a semidirect product,
$$0\to\hbox{Translations}\to\hbox{Poincar\'e}\to\hbox{Lorentz}\to 0\; ,\eek$$
and an analogy is apparent.  The connection is still closer, because there is
a unique four-dimensional normal abelian subgroup of ${\rm BMS}$ which may be
identified with the translations:
$${\rm Translations}\cong\{ \alpha\in{\rm Supertranslations}\mid\eth ^2\alpha
  =0\}\; .\eek$$
However, there is no invariant sense to a ``translation-free
supertranslation.''  Therefore there is no canonical subgroup of ${\rm BMS}$
to identify with Poincar\'e.

We shall be concerned with the action of the BMS group on the Bondi shear.  We
noted above that this shear is the shear of the $u=$ constant cuts, and so
depends on the choice of Bondi system.  In order to have a well-defined
action, then, we must specify whether the active motion is accompanied by a
passive change in Bondi system, and whether the shear is regarded as a
function on the abstract manifold $\scrip$ or a function of the coordinates
$(u,\theta ,\varphi )$.  We shall keep the Bondi system the same, and regard
the shear as a function on $\scrip$.  Then if $\phi :\scrip\to \scrip$ is the
diffeomorphism generated by the supertranslation $\alpha$, the action is
$$\sigma \mapsto\sigma\circ\phi ^{-1}+\eth ^2\alpha\;
.\eek$$\xdef\acmot{\the\EEK}%
Notice that it transforms like an ordinary function precisely under the
translations.
This may be contrasted with the result of a passive supertranslation by
$\alpha$, which action is
$$\sigma\mapsto\sigma +\eth ^2\alpha\; .\eek$$\xdef\pasmot{\the\EEK}%

\titleb{Reduction of a Phase Space by Gauge Motions}

We wish to call to mind some aspects of the reduction of a phase space, and to
fix some terminology.  Accordingly, in this sketch, we ignore all technical
difficulties (subtleties in infinite dimensions, whether quotients are
manifolds, etc.).

Let us recall the Bergmann-Dirac theory of first-class constraints (Bergmann
\& Goldberg 1955, Dirac 1964).  Let $(\Gamma ,\omega )$ be a phase space, and
suppose $\{ C_\alpha\mid\alpha\in A\}$ is a family of constraints closed under
Poisson brackets.  The symplectic form on the constraint submanifold $\Gamma
_{\rm constraint}=\{ p\in\Gamma\mid C_\alpha (p)=0\}$ is degenerate, and this
degeneracy is compensated by introducing an equivalence relation.  Let
$V_\alpha$ be the Hamiltonian vector field conjugate to $C_\alpha$:
$$dC_\alpha =\omega (\cdot, V_\alpha )\; ;\eek$$
and let ${\cal V}$ be the Lie algebra generated by these vector fields.  Then
two points in $\Gamma _{\rm constraint}$ are identified if there is a flow of
a vector field in ${\cal V}$ taking the first to the second.  The reduction of
the phase space thus has two parts:  restriction to $\Gamma _{\rm
constraint}$, and passage to a quotient.

We may run this backwards.\fonote{Jan Segert points out this is the reduction
of Marsden \& Weinstein (1974).}  Suppose a connected abelian (for simplicity)
gauge group acts on $(\Gamma ,\omega )$.  (That is, the action is a
representation of the gauge group by symplectomorphisms.)  Let $A$ be the Lie
algebra of the group, and let $V_\alpha$ be the Hamiltonian vector field
generated by $\alpha\in A$.  Furthermore, let us suppose we can find a moment
map $C:\Gamma\to A^*$, so that $dC_\alpha =\omega (\cdot ,V_\alpha )$.  Then
the
level surfaces $\Gamma _c=\{ p\in\Gamma\mid C_\alpha (p)=c_\alpha\}$ for $c\in
A^*$ play the part of the constraint manifold in the discussion above.  From
this point of view, however, there is no reason for preferring one level
surface
to another.  Therefore each such level surface, modulo the gauge action, forms
a reduced phase space.  We may think of the collection of these quotients, as
the level surface varies, as a foliation of a space whose leaves are the
reduced phase spaces.

\it We shall use the terms \rm  leaf \it and \rm foliation \it only to refer to
the possibility of choosing different level surfaces, \rm not to an orbit or
the space of orbits of the gauge action.  We may speak of a foliation in this
sense whether or not we have passed to the quotient by the gauge action.

It is the presence of such foliations that is most important in certain
radiation problems.  Passing to the quotient by the gauge action may or may not
be convenient (and in the gravitational case is definitely undesirable).
However, the foliation enters the theory by restricting the admissible vectors
to those tangent to the leaves.  This restriction turns out to be crucial in
developing a satisfactory theory of the angular momentum of radiation.

We shall show below that the radiative modes are actually fibres in certain
phase spaces, so let us see what the effects of such structure are.  Let
$(\Gamma _{\rm total},\omega )$ be a phase space; suppose there is a fibration
$$\matrix{\Gamma&\to&\Gamma _{\rm total}\cr &&\downarrow\cr
  &&\Gamma _{\rm base}\cr}\eek$$
so that $\omega$ restricts to a non-degenerate form on each fibre.  Suppose
further that the gauge action respects the fibration, that is, there is a pair
of actions $A\times \Gamma _{\rm total}\to \Gamma _{\rm total}$,
$A\times\Gamma _{\rm base}\to\Gamma _{\rm base}$ such that the diagram
$$\matrix{A\times\Gamma _{\rm total}&\to&\Gamma _{\rm total}\cr
  \downarrow&&\downarrow\cr
  A\times\Gamma _{\rm base}&\to&\Gamma _{\rm base}\cr}\eek$$
commutes.  Let $K:\Gamma _{\rm total}\to A^*$ be the moment map.

We can regard each fibre as a phase space, so suppose we have found a
moment map $C:\Gamma\to A^*$ (depending smoothly on the base point).  Then
$$\langle W,dC_\alpha\rangle =\omega (W,V_\alpha )=\langle W,dK_\alpha\rangle
\eek$$
for $W$ tangent to $\Gamma$.  Therefore we have
$$C=K+D\eek$$
for some $D$ depending only on the projection to $\Gamma _{\rm base}$.  Thus
any
level surface of the total moment map will intersect each fibre in a level
surface of the fibrewise moment map, and conversely.  It will be enough, then,
to construct the fibrewise moment maps.

\titlea{Electromagnetism}

In this Section, we give a treatment of electromagnetism on Minkowski space,
analogous to that of General Relativity below.  The case of electromagnetism is
somewhat simpler than gravity, because the gauge freedom acts ``vertically,''
in a bundle over space-time, and because the linearity of the theory allows a
cleaner isolation of the radiation field.  Still, the results here will be
useful in interpreting those for gravity.
Our emphasis here will be conceptual, and we shall give no proofs.  (These can
easily be adapted from the gravitational case, below.)  These results build on
earlier work of  Ashtekar \& Narain (1981) and Sparling, Newman and coworkers
(e.g. Chakravarty et al. 1986).

The electromagnetic potential will be denoted ${\widehat\Phi}_a$.  (Quantities
with hats refer to the physical space-time; conformally rescaled quantities
are bareheaded.)  The field strength is ${\widehat
F}_{ab}=2{\widehat\nabla}_{[a}{\widehat\Phi}_{b]}$.  We shall suppose there are
no sources, and that the field extends smoothly to $\scrip$ with the usual
peeling behavior.   We may take $\Phi _a={\widehat\Phi}_a$ to be smooth at
$\scrip$.  It is convenient to choose the gauge so that $n^a\Phi _a=0$ at
$\scrip$, where $n^a$ is a
null vector tangent to the generators of $\scrip$.  Then the most important
quantities are the rescaled components
$$\Phi =\lim _{r\to\infty} r{\widehat m}^a\Phi _a\hbox{ and }\phi _2
  =\lim _{r\to\infty}r{\widehat F}_{ab}{\widehat{\overline m}}^a
    {\widehat n}^b\; ,\eek$$
which are smooth on $\scrip$.  These are related by
$\phi _2=-\partial _u \overline\Phi$.
Note that the gauge freedom remaining at $\scrip$ is $\Phi\mapsto\Phi
+\eth\gamma$, where $\gamma$ is a smooth real-valued function of angle only.

The function $\phi _2$ can be thought of as the profile of the emitted
radiation.  It very nearly determines the electromagnetic field
throughout space-time, assuming no sources are present.  Indeed, if we consider
electromagnetic fields which vanish in some neighborhood of future timelike
infinity, then the field is determined throughout space-time by $\phi _2$.
More generally, if Cauchy data for the field are given on a spacelike
hyperboloidal initial surface $\Sigma$ meeting $\scrip$ at $u=u_0$, and if
$\phi _2$ is specified on $\scrip$ to the past of $u_0$ (and if $\phi _2\bigr|
_{u=u_0}$ agrees with the radiation field at $u_0$ which would be deduced from
the Cauchy data), then the Maxwell field is determined throughout space-time.
Since $\Phi$ determines $\phi _2$, knowledge of $\Phi$ for $u\leq u_0$ can
replace knowledge of $\phi _2$ in this argument.

The phase space may thus be constructed from two sorts of data:  the usual
Cauchy data (with suitable asymptotics) on $\Sigma$; and the field $\Phi$ on
$\cN =\{ p\in\scrip\mid u< u_0\}$.  We shall call the Cauchy data on $\Sigma$
the \it final states, \rm and the field $\Phi$ on $\cN$ the \it radiative
modes.  \rm Since $\Phi$ must agree at $Z$ with what can be deduced from the
final state, the phase space has a bundle structure:  the radiative modes
fibre over the final states.
We shall give a fibrewise analysis.

A formal construction of a fibre of the phase space is as follows.  Fix a
gauge on $\scrip$ as above, and fix a final state.  From this, we may work out
the field at $u=u_0$, and all its $u$--derivatives there.
For simplicity, assume
the field is gauge to all orders at $u=u_0$.  Let $\Gamma$ be the space of
fields $\Phi$ of type $\{ 1,-1\}$ on $\cN$ which are smooth, agree at $u_0$,
together with all their derivatives, with the values inferred from the data on
$\Sigma$, and are pure gauge sufficiently far in the past.  There is a natural
topology and smooth structure on $\Gamma$.  The sym\-plectic form is $$\omega
(f_1,f_2)=(4\pi )^{-1}\int _{u\leq u_0}
  \bigl[ (\partial _uf_1){\overline f}_2 -f_1\partial _u{\overline f}_2\bigr]
  \, dud{\cal S}+\hbox{conjugate}\eek$$
(Ashtekar \& Streubel 1981).  It is closed and weakly non-degenerate.

Now let us consider the effect of a gauge change.  The admissible changes are
of the form $\Phi\mapsto\Phi +\eth\gamma$ on $\scrip$, where $\gamma$ is a
smooth function on the sphere.\fonote{Strictly speaking, there is not a
well-defined action by such functions on the space of data on $\Sigma$, since
there are many gauge changes on the interior of $\Sigma$ with the same
asymptotic behavior.  Thus a correct treatment has the space of $\gamma$'s as
a quotient of a larger gauge freedom.  The result is the same.}  We let
$V_\gamma$ be the vector field on $\Gamma$ generating this motion.  Then one
can check that
$$\omega (\cdot ,V_\gamma )=dC_\gamma\; ,\eek$$
where
$$C_\gamma =(4\pi )^{-1}\oint [\![\Phi ]\!]\eth '\gamma \, d{\cal S}\Bigr|
_{u=-\infty}^{u_0}
  +\hbox{ conjugate}\; .\eek$$
As $\gamma$ varies, this detects precisely the electric part of
the gauge-invariant jump
$$[\! [\Phi ]\! ]=\Phi\Bigr| _{u=u_0}-\Phi \Bigr| _{u=-\infty}\; .\eek$$
Therefore the leaves of the foliation of the phase space
are labeled by the electric parts of these jumps.\fonote{It should be
emphasized
that the term ``electric'' is used in the sense of spin- and boost-weighted
functions, and is not obviously related to the electric part of the
electromagnetic field.  However, it can be shown that $C_\gamma =(2\pi )^{-1}
\oint [\! [\Re \phi _1]\! ] \gamma\, d{\cal S}$ where $\Re\phi _1$ is a certain
component of the electric field in the Bondi system.}

Now let us consider fields for which $\Phi$ is pure gauge (i.e., purely
electric) at $u_0$.  Those with $[\! [\Phi ]\! ]=0$ are said to lie in the \it
classical \rm sector; the others are in \it infrared \rm sectors, with the
sector labeled by the jump.   Such sectors play an important role in Quantum
Electrodynamics, in the construction of the Hilbert space of
out-states.\fonote{In our notation, the usual construction of infrared sectors
corresponds to the case $u_0=+\infty$.  It seems likely, in view of the
considerations of well-posedness discussed above, that some of the difficulties
in the usual construction may be removed by delaying the limit $u_0\to +\infty$
to a later stage in the analysis.  Compare Jauch \& Rohrlich (1976).}
Essentially, the sector is determined by the out-state of the charged
particles, and a Hilbert space of acceptable states of the electromagnetic
field, given these data, is constructed from fields in the sector.  It was
rather a surprise to find that the electromagnetic field was so restricted
(Ashtekar \& Narain 1981).  However, we can now see that this is foreshadowed
by the foliation of the classical phase space:

\begtheorem {3.1}
Fix a final state for which $\Phi$ is pure gauge at $Z$.  Let $C_\gamma$ be
the moment map for the gauge freedom on $\scrip$.  Then the classical
sector is the zero-set of this moment map, and the infrared sectors are the
level sets for non-zero values.
\endtheorem

\titlea{The Space-Times}

The phase space for the outoing radiative modes in General Relativity will be
constructed as a certain function space, whose elements represent data for an
initial-value problem for Einstein's equations.  In this Section, we discuss
the space-times which are the solutions of these problems.
We first review how the data determine the solutions locally, and
then give a rapid treatment of the elementary aspects of the global
theory.

\begfig 6 cm
\figure{1}{The initial-data surface $\cH =\Sigma\cup\cN$ for the space-time.
(Cross-section.)}
\endfig

\titleb{Local Results}

It will be clearest first to describe the sort of solutions sought, and then to
characterize the data.

We shall want a vacuum space-time $\whst$, oriented and time-oriented,
with $\widehat M$ embedding as the interior of a manifold with boundary $M$,
and the following properties:
\medskip
\item{(a)} There exists a smooth function $\Omega$ on $M$, positive on
$\widehat M$ and zero on $\scrip =\partial M$, with $\nabla _a\Omega$ nowhere
zero on $\scrip$;

\item{(b)} The metric $g_{ab}=\Omega ^2{\widehat g}_{ab}$ extends smoothly
to a non-degenerate metric on $M$;

\item{(c)} $\scrip$ is a null hypersurface
diffeomorphic to $\R \times S^2$, with the ``$\R$'' factors being the null
generators;

\item{(d)} The points on $\scrip$ are the future end-points of null
geodesics in $\whst$;

\item{(e)} There exists a partial Cauchy surface $\whSig$ in $\whst$ whose
closure $\Sigma$ in $M$ is compact and meets $\scrip$ transversely in a cut
$Z$ (i.e., a section of the fibration $\scrip\to S^2$);

\item{(f)} A Bondi retarded time coordinate attains arbitrily negative
values on each generator of $\scrip$.
\medskip
\noindent Requirements (a)-(d) are standard for a future null infinity in
radiation problems.  Property (e) will turn out to ensure that the asymptotic
initial-value problem is well-posed:  data will be given on $\Sigma$ and the
portion $\cN$ of $\scrip$ to the past of $Z$.  Property (f) is also a standard
requirement; it could be weakened, however, without much change.

Now let us explain how such a space-time is determined by data on $\cH
=\Sigma\cup\cN$.

Let $\Sigma$ be an oriented three-manifold whose boundary $Z$ is diffeomorphic
to $S^2$; let $\whSig$ be the interior of $\Sigma$.  Let regular hyperboloidal
initial data for Einstein's equations, in the sense of Friedrich (1983), be
given on $\Sigma$.  (Actually, we consider the time-reverse of Friedrich's
case.
Also, Friedrich took $\Sigma$ to be the ball, but this is unnecessary here.)
These data determine a space-time for which a portion of $\scrip$ to the future
of $Z$ exists.  The past Cauchy horizon $N$ of $\Sigma$ in this space-time is
(near $Z$) a smooth null hypersurface and the metric and its derivatives on
this hypersurface attain limits which determine smooth asymptotic
characteristic data.  Properties (a)-(e) hold.

Fix a Bondi system at $\scrip$ in which $Z$ corresponds to $u=0$.  We may then
work out, from the space-time above, the Bondi shear and all its
$u$-derivatives at $Z$.  Let $\cN =(-\infty ,0]\times S^2$, and
let $\sigma$ be any smooth extension of the Bondi
shear to $\cN$.  Then a theorem of Rendall (1990) shows that the
doubly-characteristic initial-value problem on $\cN\cup N$ is
well-posed locally relative to $Z$, and so there is a locally unique
space-time inducing these data in a neighborhood of $Z$ as well.

The space-times under consideration are then determined (at least in a
neighborhood of $\cH$) by the regular hyperboloidal initial data on $\Sigma$
and the shear on $\cN$.  Therefore such data ought to form the manifold of the
phase space.  Since the admissible shears depend on the hyperboloidal data, we
see that the phase space has a bundle structure, with the spaces of allowed
shears fibering over the hyperboloidal data.  We shall refer to the data on
$\Sigma$ as the \it final state, \rm and the shears compatible with a final
state as the \it radiation data.\rm\/  We are
here concerned exclusively with a fibrewise analysis:  we fix the final state,
and consider the possible radiation data compatible with it.

At present, it is not known under what conditions a final state will admit a
radiative datum for which a space-time exists in a neighborhood of $\cH$,
that is, for which condition (f) holds.  However, for local analysis on the
phase space, it is not necessary to know exactly which data are admissible.  We
simply need to know that sufficiently small perturbations of a space-time
satisfying (a)-(f) will also satisfy (a)-(f), and that the map from data to
space-times is smooth.  These are guaranteed by stability results of Rendall
(1992).

\titleb{Developments}

We now show that the data under consideration have well-defined maximal
globally hyperbolic extensions.  The argument is quite similar to that for the
standard Cauchy problem, and so a rapid treatment will be given.  However, the
force of the present results is somewhat stronger from that for the standard
Cauchy problem.  In the standard case, a development of the data by definition
contains the entire initial surface as a Cauchy surface.  Once a local
existence and uniqueness theorem is established, this is no loss of generality,
since the finite speed of propagation guarantees the existence of such
developments.  In the characteristic case, on the other hand, the local
existence and uniqueness results do {\it not}\/ guarantee the existence of
developments including the entire initial surface.  Therefore one must
consider developments with different domains on the initial surface, and the
maximality one seeks is in part a maximality with respect to such domains.

This result is somewhat stronger than what will be needed below, since in the
sequel we shall simply posit that the solution extends to the whole of the
initial-data surface.  However, we recall that our aim is to give a
construction that is robust, and for this reason the result is of interest
here.

In what follows, we assume a familiarity with the differential topology of
space-times, as described in Penrose (1972).  A few comments are
necessary, since the standard expositions of the theory apply to finite
space-time, and we shall have to consider the boundary as well.

The notions of chronological and causal precedence
are conformally
invariant and so have an interpretation in $\st$ independent of the factor
$\Omega$ used indefining the conformal completion.  Thus the chronological
and causal futures and pasts $I^+$, $J^+$, $I^-$, $J^-$ have evident
meanings.  The concept of an \it achronal set \rm is also well-defined.  Then
the domains of dependence of an achronal set are also defined.  We shall soon
show that $\cH$ is an achronal set in the solutions of interest; then we shall
be interested in the domains
$$\eqalign{D^\pm (\cH )=\{ x\in\widehat M \mid
  &\hbox{ every timelike curve in }M\hbox{ through }x \hfill\cr
  &\hbox{ which is past- (future)-endless in }
  \widehat M \hbox{ meets }\cH\}\cr}$$
and $D(\cH )=D^+(\cH )\cup D^-(\cH )$.

In what follows, when condition (f) is not assumed, the sets $\cH$ and $\cN$
are
understood to include only the portion of $\scrip$ which exists.

\beglemma {}
Let $\whst$ be a time-oriented space-time satisfying (a)-(e), and suppose
$\cH$ is achronal.  Then ${\cal M}=\Int D(\cH )$ is globally hyperbolic.
Also $\whSig\subset {\cal M}$, and every point on $\cN$ is the endpoint of a
null geodesic on ${\cal M}$.
\endlemma

\begproof\/
That ${\cal M}$ is strongly causal follows from Lemma~4.16 in Penrose~(1972).
That $J^+(p)\cap J^-(q)$ is compact for $p,q\in {\cal M}$ follows from
Proposition~5.20 and Theorem~6.5 there.  The remaining assertions are
elementary.
\qed
\endproof

\begdefinition {}
Let a final state on $\Sigma$ and a radiation datum compatible with it on
$\cN$ be given.  A {\rm development} of these data is an oriented,
time-oriented space-time $\whst$ for which properties (a)-(e) hold (and
inducing the correct data on $\Sigma$ and that portion of $\scrip$ which
exists), such that $\cH$ is $g_{ab}$-achronal and
${\widehat M}=D(\cH )$.
\enddefinition

\beglemma {}
For any final state and radiation datum compatible with it, a development
exists.
\endlemma

\begproof\/
The results of Friedrich together with standard results on the Cauchy
problem show that there is a maximal globally hyperbolic development of the
data on $\whSig$ for which a portion of $\scrip$ at and to the future of $Z$
exists, and for which the past Cauchy horizon is (near $Z$) a smooth null
hypersurface meeting $\scrip$ transversely in $Z$.  Moreover the field attains
smooth limits on this hypersurface, which are the correct characteristic
initial data.  Let us denote the manifold of this space-time by $M_{\rm F}$.

Rendall's theorem shows that the initial-value problem with the
data on the past horizon and those on $\cN$ is well-posed (locally at $Z$).
Let $M_{\rm R}$ be a space-time produced by this theorem, with $S$ the strip of
the past Cauchy horizon of $\whSig$ serving as one of the characteristic data
surfaces.  (We do not include $Z$ in $S$.)

Now $M_{\rm F}\cup S\cup M_{\rm R}$ is an oriented, time-oriented space-time
satisfying (a)-(e).  Also no point in $M_{\rm R}$ is in the future of any point
in $M_{\rm F}$.  It follows that $\cH$ is achronal in this space-time.
The remaining claims will follow from the previous lemma.
\qed
\endproof

\begtheorem {4.1}
There exists a unique development which is maximal with respect to inclusion.
\endtheorem

\noindent (Here ``unique'' and ``inclusion'' are to be understood in the sense
of the natural isomorphism classes of developments.)

\begproof\/
Fix a Bondi coordinate system on $\scrip$ in which the cut $Z$ is given by
$u=0$.  We shall consider developments meeting $\scrip$ in different sets, and
we shall want to compare their sets $\cN$ of points on $\scrip$ to the past of
$Z$.  We begin by examining the forms of these sets.

Condition (c), that $\cN$ be diffeomorphic to $\R\times S^2$ with null
generators being the ``$\R$'' factors, may be expressed more formally by saying
that there is a diffeomorphism $\R\times S^2\to\cN$
of the form
$$\bigl( x,(\theta ,\varphi )\bigr)\mapsto
    \bigl( u(x,\theta ,\varphi ),(\theta ,\varphi )\bigr)\; .$$
The function $\alpha :S^2\to [-\infty ,0 )$ given by
$\alpha (\theta ,\varphi )=\inf _xu(x,\theta ,\varphi )$
is, as an infimum of continuous functions, upper semicontinuous.  Conversely,
it is not hard to show that for any upper semicontinuous function $\alpha
:S^2\to [-\infty ,0 )$, the set
$$\cN (\alpha )=
  \{ (u,\theta ,\varphi )\in\scrip\mid 0>u>\alpha (\theta ,\phi )\}$$
satisfies condition (c).  (We do not claim that there is a development for
every such set, however.)

If a development exists corresponding to a function $\alpha$, then there is a
unique maximal globally hyperbolic development corresponding to this
function.  This argument follows that for the standard Cauchy problem, cf.
Hawking \& Ellis (1973).  (One must be a little careful because of
the freedom in choosing the conformal factor.  However, since there are
geometric ways of fixing this in a neighborhood of $\scrip$, there is no real
difficulty.)

Now let $A$ be the set of $\alpha$'s achievable from the given data.  If
$\alpha _1,\alpha _2\in A$ with $\alpha _1\leq\alpha _2$ (for all $\theta$,
$\varphi$), then the maximal globally hyperbolic development
${\whst}_1$ corresponding to $\alpha _1$ must include the one ${\whst}_2$
corresponding to $\alpha _2$.  To see this, note that we can glue to
${\whst}_2$ a portion of ${\whst}_1$ on the region extending inwards from
$\cN (\alpha _1)-\cN (\alpha _2)$  to get a development, which must then
be included in ${\whst}_1$.

Let $\underline\alpha (\theta ,\varphi )=\inf _{\alpha\in A}\alpha (\theta
,\varphi )$.  Since $\cN (\underline\alpha )=\bigcup _{\alpha\in A}\cN (\alpha
)$, we can construct a solution to the initial-value problem in a neighborhood
of $\Sigma\cup\cN (\underline\alpha )$ by patching together solutions in
neighborhoods of the sets $\Sigma\cup \cN (\alpha )$.  From this we can
construct a development corresponding to $\underline\alpha$, and so there is a
unique maximal globally hyperbolic development $\whst$ corresponding to this
function.  By the argument of the previous paragraph, it is maximal among the
developments for arbitrary $\alpha\in A$.
\qed
\endproof

\titlea{The Manifold Underlying the Phase Space}

As discussed above, the full phase space fibres over the space of final
states, and we are concerned with the structure of a fixed fibre.
Physically, we have in mind space-times which are quiescent (as far as
outgoing radiation is concerned) in the far past and again in the far future.
For technical simplicity, we shall only consider the case where $\sigma$ is
electric at $Z$ and its positive derivatives with respect to $u$ vanish there.
(This leads to a particularly clean isolation of the radiative degrees of
freedo
from the internal ones.)  For quiescence in the far past, we
require that for sufficiently negative $u$ the shear is independent of $u$ and
is purely electric.  Then the allowed space $\Gamma$ of $\sigma$'s will be the
set of those agreeing at $Z$, together with all their $u$--derivatives, with
that inferred from the hyperboloidal data on $\Sigma$, and which are pure gauge
(i.e., $\eth ^2\alpha$ for some real function $\alpha$ independent of $u$) in
the far past.
It is in order to recall here that not every such datum may correspond to a
space--time.  (See the last paragraph in Section 4.1.)  Therefore the physical
phase space is really some open set in $\Gamma$.

In order to have a uniform treatment of the gauge freedom, it will be
convenient not to fix the supertranslational freedom at first.  Since we are
considering shears which are pure gauge at $Z$, this will have the effect of
allowing arbitrary such shears.

Fix a diffeomorphism $\cN\to (-\infty ,0]\times S^2$.  Let $\Gamma$ be the set
of all smooth functions $\sigma$ of type $\{ 3,-1\}$ on ${\cal N}$ whose
$u$-derivatives vanish outside a compact set, which are electric at $u=0$
and for $u$ sufficiently negative.  The most convenient way to topologize
$\Gamma$ is as follows.  Choose any function $f$ (of type $\{ 0,0\}$) on ${\cal
N}$ which is smooth, identically $-1$ for sufficiently negative $u$ and
identically $+1$ close to $u=0$.  The map
$$C^\infty (S^2,\{ 3,-1\} )_{\rm electric}\oplus C^\infty (S^2,\{ 3,-1\} )_{\rm
electric}
  \oplus C_0^\infty (\cN ,\{3,-1\})  \to
\Gamma\eek$$\xdef\charry{{$\the\EEK$}}%
given by
$$(\alpha ,\beta ,\gamma )\mapsto \alpha f+\beta +\gamma\eek$$
is an isomorphism of (algebraic) vector spaces.  (Here the electric elements of
$C^\infty (S^2,\{ 3,-1\})$, as the kernel of a certain differential operator,
form a Fr\'echet space.  The subscript $0$ denotes compact supports.)
Topologize the right-hand side so that this is a homeomorphism.

\beglemma {} The topology so defined is independent of the choice of $f$.
\endlemma

\begproof\/
Let $f'$ be a second such function.  Then we suppose that we have
$$\alpha f+\beta +\gamma =\sigma =\alpha 'f'+\beta '+\gamma '\; .\eek$$
Considering sufficiently small $u$, we see that we must have $-\alpha +\beta
=-\alpha '+\beta '$; similarly, for sufficiently large $u$, we must have
$\alpha +\beta =\alpha '+\beta '$.  Thus $\alpha =\alpha '$ and $\beta =\beta
'$, and so $\gamma '=\gamma +\alpha (f-f')$.  It follows that the transition
function
$$\eqalign{
C^\infty &(S^2,\{ 3,-1\} )_{\rm electric}\oplus C^\infty (S^2,\{ 3,-1\} )_{\rm
electric}
  \oplus C_0^\infty (\cN ,\{ 3,-1\})\hfill\cr
  &\to C^\infty (S^2,\{ 3,-1\} )_{\rm electric}\oplus C^\infty (S^2,\{ 3,-1\}
)_{\rm electric}
  \oplus C_0^\infty (\cN ,\{ 3,-1\})\cr}$$
given by
$$(\alpha ,\beta ,\gamma )\mapsto (\alpha ',\beta ',\gamma ')=
  (\alpha ,\beta ,\gamma +\alpha (f-f'))$$
is a homeomorphism.
\qed
\endproof

Thus the inverses of the maps (\charry{}), as $f$ varies, form an atlas for
$\Gamma$.  The transition function identified in the proof above is
smooth, so we have given $\Gamma$ the structure of a smooth
manifold.

The following are immediate:  (a) the manifold is
modeled on a sequentially complete locally convex topological vector
space; (b) this vector space is reflexive and its dual may be identifed with a
suitable space of distributions; (c) the sequence
$$0\to\hbox{Gauge Freedom}\to\hbox{Unreduced
Data}\to\hbox{Bondi News Functions}
  \to 0\eek$$
given by
$$\matrix{
0&\to&C^\infty (S^2,\{ 3,-1\} )_{\rm electric}&\to &\Gamma&\to &C_0^\infty (\cN
,\{ 2,0\} )
    &\to &0\cr
           &        &\beta &\mapsto &(0,\beta  ,0)&  &
 &  & \cr
           &        &      &        &(\alpha ,\beta ,\gamma ) &\mapsto &
\alpha\partial _uf+\partial _u\gamma &  & \cr
          }\eek$$
is an exact sequence of topological vector spaces; (d) the projections from
$\Gamma$ to $C^\infty (S^2,\{ 3,-1\})_{\rm electric}$ giving the future and
past limits of the shear  are smooth.

Property (c) implies that the smooth structure of $\Gamma$ is invariant
under passive supertranslations.  It is not hard to show too that the smooth
structure is invariant under passive Lorentz motions, so that it is invariant
under all passive BMS motions.  For active motions, we must
consider the family of $\Gamma$'s as the cut $Z$ varies; the naturality of our
construction guarantees the appropriate covariance.

Finally, we remark that the topology on $\Gamma$ is somewhat finer than that
needed in the theorems of Rendall (or in what follows).

\titlea{The Symplectic Form}

Ideally, one would like to identify a symplectic form on the phase space we
have
constructed by starting with the usual symplectic form got by integrating
linearized solutions over Cauchy surfaces (Arnowitt, Deser \& Misner 1962,
Regge \& Teitelboim 1974, Chernoff \& Marsden 1974), and working out what this
gives in terms of our data.  Formally, the way to proceed is clear.  Since the
three-form which is integrated is closed, one should deform the Cauchy surface
to the M-shaped surface $\cH$.  This approach was outlined in Ashtekar \&
Magnon-Ashtekar (1982).  In our notation, the symplectic form they identified
is given by
$$\omega (s_1,s_2)=(8\pi G)^{-1}\int _{\cN}\bigl[ {\dot
s}_1\overline{s_2}-
   \dot{\overline{s}}_2 s_1\bigr] \, dud{\cal S}
  +\hbox{ conjugate}\; .\eek$$

Since the surfaces involved are not compact, some justification of the
invariance of the integral under deformations is necessary.  The difficulty
occurs at $u=-\infty$.  The difference between integrating over the M-shaped
surface and the Cauchy surface amounts to what may be called a leakage
integral (at $u=-\infty$, which presumably corresponds to spacelike
infinity).  A comprehensive treatment of such leakages would require stronger
asymptotic estimates than are currently established.\fonote{Roughly speaking,
one would want to show that the asymptotic expansion of the metric held
uniformly as $u\to -\infty$ to $O(1/r)$.  The relation between null and
spacelike infinity is a major outstanding problem in the existence theory for
Einstein's equations.}  Therefore we shall give a proof of the correctness of
the symplectic form by less direct means.  This proof has the advantage of
avoiding all gauge problems.

We first verify that $\omega$ \it is \rm a symplectic form.

\begproposition {}
The form $\omega$ is smooth, closed and weakly non-degenerate.
\endproposition

\begproof\/
It is smooth and closed because it is constant in the chart.  Integrating
by parts, we have
$$\omega (s_1,s_2)=(4\pi G)^{-1}\int _{\cN} {\dot s}_1\overline{s_2}
  \, dud{\cal S}-(8\pi G)^{-1}\oint \overline{s_2}s_1\, d{\cal S}
  +\hbox{conjugate}\; ,$$
where $\oint$ denotes the difference between the boundary integrals at $u=0$
and $u=-\infty$.  Suppose, for some fixed $s_1$, this vanishes for all
$s_2$.  By taking $s_2$ to be a bump function in $u$ times an arbitrary
angular function (with values in $\{ 3,-1\})$, we have ${\dot s}_1=0$.
However, then we have
$$\omega (s_1,s_2)=-(8\pi G)^{-1}\oint [\! [\overline{s_2} ]\! ]s_1
  \, d{\cal S}+\hbox{conjugate}\; ,$$
where $\oint$ now is the usual angular integral.  This implies that the
electric part of $s_1$ is zero.  However, since $s_1$ is constant, it must be
purely electric.
\qed
\endproof

\begfig 8 cm
\figure{2}{Construction of the Cauchy surface in Theorem~6.1.  A cross-section
is shown.}
\endfig

\begtheorem {6.1}
Let $\whst$ be the maximal globally hyperbolic development of data on $\cH$,
and let $s_1$, $s_2$ be tangent vectors at the corresponding point in $\Gamma$.
Let $h_1$, $h_2$ be linearized solutions to Einstein's equations inducing the
data $s_1$, $s_2$ on $\cN$ (and pure gauge on $\Sigma$).  Then there is a
Cauchy
surface $S$ in $\whst$ such that $\omega (s_1,s_2)=\omega _S(h_1,h_2)$, where
$\omega _S(h_1,h_2)$ is the usual $3+1$ symplectic form determined by
integrating over $S$. \endtheorem

\begproof\/
We first show that $\omega _{\whSig}(h_1,h_2)=0$.
Let $h_{1ab}={\widehat\nabla}_{a}\xi _{1b}+{\widehat\nabla}_{b}\xi _{1a}$, and
similarly for $h_2$ and $\xi _2$.  Let $\alpha _1$, $\alpha _2$ be the
supertranslations induced by $\xi _1$, $\xi _2$ at $Z$.  Then
$\omega _{\whSig}(h_1,h_2)=\langle h_1,dP_2\rangle$ where $P_2$ is the
supermomentum at $Z$ determined by $\alpha _2$ (and $\langle
\cdot,\cdot \rangle$ is pairing in the $3+1$ phase space).  Then
$\langle h_1,dP_2\rangle$ is the change in this supermomentum under a
supertranslation by $\alpha _1$.  However, the vanishing of the derivatives of
$\sigma$ at $Z$ implies this is zero.

Assume the Bondi system has been chosen so that $Z$ is given by $u=0$.  Fix
the coordinate $\rho =\Omega$ as in Penrose \& Rindler (1986), section
9.8.  (The relevant effect of this is to ensure that, for any fixed $u$, the
surfaces of constant $\rho$ are space-like for small enough positive $\rho$.)
For negative integers $n$, consider the null hypersurface extending
orthogonally inwards from $u=n$.
We choose a smooth portion $C_n$ of this
hypersurface given by $\rho\leq c_n$ for some positive $c_n$.  (We shall place
restrictions on $c_n$ in what follows.)

Now assume $n$ is so negative, say $n\leq N$, that $s_1$, $s_2$ are pure gauge
for $u\leq n$.  Then $\omega (s_1,s_2)=\omega _{K_n}(h_1,h_2)$, where
$$K_n=\{ p\in\whSig\mid \rho (p)\geq c_n\}\cup\{ p\in\widehat{M}\mid
  \rho (p)=c_n\hbox{ and } n\leq u(p)\leq 0\}\cup C_n\; .$$

For each $n\leq N$, consider the
contribution to the integral $\omega _{K_n}(h_1,h_2)$ from $C_n$.  By choosing
$c_n$ small enough, we may make this contribution arbitrarily small; indeed,
we may make the integral of the absolute value of the integrand arbitrarily
small.  Choose $c_n$ small enough so that this latter is $<1/|n|$.

Now construct $S$ by starting with
$$\{ p\in\whSig\mid \rho (p)\geq c_N\}$$
and for $n\leq N$ alternately joining the surfaces
$$\{ p\in\widehat M \mid \rho (p)=c_n\, ,\ n-1\leq u\leq n\}$$
with
$$\{ p\in\widehat M\mid p\in C_{n-1}\hbox{ and }c_{n-1}\leq\rho (p)\leq c_n\}\;
.$$
This surface is by construction everywhere spacelike or null, and $\omega
(s_1,s_2)=\omega _S(h_1,h_2)$.  We must show it is a Cauchy surface.

A timelike curve starting from a point on $S-\Sigma$ will have $\rho$ strictly
decreasing and $u$ strictly increasing.  Since $\rho$ is a monotonically
increasing function of $u$ on $S-\Sigma$, this curve cannot meet $S-\Sigma$ at
a second point.  If it meets $\Sigma$, it must do so on the set of points with
$\rho <c_N$, that is, on $\Sigma -S$.  So there is no timelike curve from
$S-\Sigma$ to $S$.  Since we already know $\Sigma$ is achronal, the set $S$
must be achronal.

Finally, let us show that we can ensure that every null geodesic meets $S$.
We know every null geodesic meets $\cH$.  Let $Y$ be the space of null
geodesics, topologized as bundle over $\cH$.  (So the fibre is $S^2$ over a
point in $\whSig$ and $\R ^2$ over a point in $\cN$.)  Write $Y=\bigcup
_{n=1}^\infty Y_n$ where each $Y_n$ is compact and has base contained in
$$\Sigma\cup\{ p\in \cN\mid u>-n\}\; .$$
Thinking of the elements of $Y_n$ as past-directed null geodesics from
$\cH$, they must meet $K_n$ in a compact set for which $\rho$ is bounded
away from zero.
By choosing $c_n\to 0$ quickly enough as $n\to -\infty$, then, we can
ensure that every null geodesic meets $S$.
\qed
\endproof

Note that in this result the Cauchy surface may depend on $s_1$ and $s_2$; on
the other hand, no gauge restrictions were placed on $h_1$ and $h_2$ except
that they should induce the correct data on $\cH$.  These issues are related:
only if we make some choice of gauge in the interior of space-time can we
expect to control the integrals uniformly.  However, gauge-fixing is not the
only obstacle to getting some sort of uniform control; the relation between
null infinity and spatial infinity would have to be understood.

\titlea{Gauge Freedom}

The sectors for the gravitational phase space are defined by considering the
jumps $[\! [\sigma ]\! ]$, which are purely electric functions on the sphere.
The \it classical sector \rm is $[\! [\sigma ]\! ] =0$; the others are \it
infrared sectors \rm (Ashtekar 1987).  They play an important role in the
theory of angular momentum.

The interpretation of the sectors for gravity turns out to be sublty different
from that for electromagnetism.

\begtheorem {7.1}
The sectors are the level surfaces of the function $C:\Gamma\to {\rm
Supertranslations}^*$ given by
$$C_\alpha =(8\pi G)^{-1}\oint [\! [\sigma ]\! ]{\eth '}^2\alpha\, d{\cal S}
  +\hbox{\rm conjugate}\; .$$
This function descends to a moment map for the group
{\rm Supertranslations/Trans\-la\-tions} acting by
$\sigma\mapsto\sigma +\eth ^2\alpha$.
\endtheorem

\begproof\/
The first assertion follows from the fact that $C^\infty (S^2,\{ 3,-1\} )$ and
$C^\infty (S^2,\{ -3 ,1\}$ are dual, together with the surjectivity of the map
${\eth '}^2:C^\infty (S^2,\{ 1,1\} )\to C^\infty (S^2,\{ -1,3\} )$.

Certainly the action $\sigma\mapsto\sigma +\eth ^2\alpha$ descends to the
quotient group.  Let $V_\alpha$ be the vector field generating this motion.
Then we have
$$\eqalign{\omega (s,V_\alpha )
  &=(8\pi G)\int _{\cN} {\dot s}{\eth '}^2\alpha\, du d{\cal S}
    +\hbox{conjugate}\cr
  &=(8\pi G)^{-1}\oint [\! [s]\! ]{\eth '}^2\alpha\, d{\cal S}
    +\hbox{conjugate}\cr
  &=\langle s,dC_\alpha\rangle\cr}$$
\qed
\endproof

Notice that the action here is \it not \rm the standard active BMS motion,
Eq.~(\acmot{}), unless $\dot\sigma =0$.  Although formally the same as the
result of a passive BMS motion, Eq.~(\pasmot{}), we do \it not \rm accompany
this motion by a change in the Bondi parameter, so this is \it not \rm a
passive BMS motion, either.  We shall see in a subsequent paper that this
foliation, and the difference between the action here and the standard BMS
actions, goes some way to resolving an old mystery:  why approaches to defining
angular momentum at cuts of $\scrip$ by BMS motions give unphysical answers.

\acknow{It is a pleasure to thank John Beem  and Jan Segert for useful
conversations and Alan Rendall for useful electronic exchanges.}

\begref{References}{}

\ref Arnowitt, R.L., Deser, S. \& Misner, C.W. (1962):  Canonical Analysis of
General Relativity.  In:  Recent Developments in Relativity.
New York, Pergamon.

\ref Ashtekar, A. (1987):  Asymptotic quantization.  Naples, Bibliopolis.

\ref Ashtekar, A. \& Magnon--Ashtekar, A. (1982):  On the Symplectic
Structure of General Relativity, Comm. Math. Phys. \bf 86\rm , 55--68.

\ref Ashtekar, A. \& Narain, K.S. (1981): Infrared Problems and Penrose's
Null Infinity, International Conference on Mathematical Physics, Berlin.

\ref Ashtekar, A. \& Streubel, M. (1981): Symplectic Geomtery of Radiative
Modes and Conserved Quantities at Null Infinity, Proc. R. Soc. Lond.
\bf A376\rm , 585--607.

\ref Bergmann, P.G. \& Goldberg, I. (1955): Dirac Bracket Transformations
in Phase Space, Phys. Rev. \bf 98\rm , 531--538.

\ref Bondi, H. (1960): Gravitational Waves in General Relativity,
Nature \bf 186\rm , 535.

\ref Bondi, H., van der Burg, M.G.J., and Metzner, A.W.K. (1962): Gravitational
Waves in General Relativity. VII.  Waves from Axi-symmetric Isolated Systems,
Proc. R. Soc. Lond. \bf A269\rm , 21--52.

\ref Chakravarty, S., Ivancovich, J. and Newman, E.T. (1986): ``Infrared''
Maxwell fields, Gen. Rel. Grav. \bf 18\rm , 633-40.

\ref Dirac, P.A.M. (1964): Lectures on quantum mechanics.  Belfer
Graduate School of Science, Yeshiva University, New York.

\ref Friedrich, H. (1983):  Cauchy problems for the conformal vacuum field
equations in general relativity.  Commun. Math. Phys. \bf 91\rm , 444--472.

\ref Friedrich, H. (1992): Asymptotic structure of space-time.  In: A.I.~Janis
and J.R. Porter (eds.):  Recent advances in general relativity.  Birkh\"auser,
Boston-Basel-Berlin.

\ref Hawking, S.W. \& Ellis, G.F.R. (1973):  The large scale structure of
space--time.  Cambridge, University Press.

\ref Helfer, A.D. (1990):  The angular momentum of gravitational radiation.
Phys. Lett. \bf A150\rm , 342--344.

\ref Horowitz, G.T. \& Tod, K.P. (1982):  A relation between local and total
energy in general relativity.  Commun. Math. Phys. \bf 85\rm , 429--447.

\ref Jauch, J.M. \& Rohrlich, F. (1976):  The theory of photons and electrons.
New York, Springer-Verlag.

\ref Ludvigsen, M. \& Vickers, J.A.G. (1981): The positivity of the Bondi mass.
J. Phys. \bf A14\rm , L389--91.

\ref Marsden, J. \& Weinstein, A. (1974): Reduction of symplectic manifolds
with symmetry. Rep. Math. Phys. \bf 5\rm , 121--130.

\ref Moncrieff, V. (1975): Spacetime symmetries and linearization stability of
the Einstein equations.  I.  J. Math. Phys. \bf 16\rm , 493--498.

\ref Moncrieff, V. (1976): Spacetime symmetries and linearization stability of
the Einstein equations.  II.  J. Math. Phys. \bf 17\rm , 1893--1902.

\ref Newman, E.T. \& Penrose, R. (1962):  An approach to gravitational
radiation by a method spin coefficients.  J. Math. Phys. \bf 3\rm , 896--902.
(Errata \bf 4 \rm (1963) 998.)

\ref Newman, E.T. \& Penrose, R. (1966): Note on the Bondi--Metzner--Sachs
group. J. Math. Phys. \bf 7\rm , 863--70.

\ref Penrose, R. (1963): Asymptotic properties of fields and space--times.
Phys. Rev. Lett. \bf 10\rm , 66-8.

\ref Penrose, R. (1972):  Techniques of Differential Topology in Relativity.
Philadelphia, SIAM.

\ref Penrose, R. (1982):  Quasi--local mass and angular momentum in general
relativity. Proc. R. Soc. Lond. \bf A381\rm , 53--63.

\ref Penrose, R. \& Rindler, W. (1984--6): Spinors and space--time.
Cambridge, University Press.

\ref Regge, T. \& Teitelboim, C. (1974): Role of surface integrals in the
Hamiltonian formulation of general relativity.  Ann. Phys. \bf 88\rm
286--318.

\ref Rendall, A.D. (1990):  Reduction of the characteristic initial value
problem to the cauchy problem and its applications to the Einstein equations.
Proc. R. Soc. Lond. \bf A427\rm , 221--239.

\ref Rendall, A.D. (1992):  Stability in the characteristic initial value
problem.  In:  Z. Perj\'s (ed.): Relativity today.  Commack, New York, Nova
Science.

\ref Sachs, R.K. (1962a):  Gravitational waves in general relativity.  VIII.
Waves in asymptotically flat space--time.  Proc. R. Soc. Lond. \bf A270\rm ,
103--26.

\ref Sachs, R.K. (1962b):  Asymptotic symmetries in gravitational theory.
Phys. Rev. \bf 128\rm , 2851--64.

\ref Schoen, R. \& Yau, S.T. (1982):  Proof that the Bondi mass is positive.
Phys. Rev. Lett. \bf 48\rm , 369--71.

\endref

\bye